\documentclass[preprintnumbers,a4paper,twocolumn,floats,aps,nofootinbib]{revtex4}
\usepackage{graphicx}
\usepackage{amsmath,amssymb,amsfonts}
\usepackage{hyperref}
\usepackage{bbm}
\usepackage{xspace}
\usepackage{verbatim}
\usepackage{bm}
\usepackage{color}
\usepackage{cancel}
\usepackage{ulem}
\usepackage{bbm}

\begin{document}
\title{Modulated Natural Inflation}
\author{Rolf Kappl\footnote{kappl@th.physik.uni-bonn.de}, %
Hans Peter Nilles\footnote{nilles@th.physik.uni-bonn.de}, %
Martin Wolfgang Winkler\footnote{winkler@th.physik.uni-bonn.de}}

\affiliation{
\vspace{4mm}
Bethe Center for Theoretical Physics and Physikalisches Institut der Universit\"at Bonn\\
Nussallee 12, 53115 Bonn, Germany}

\begin{abstract} 
We discuss some model-independent implications of embedding
(aligned) axionic inflation in string theory. As a consequence
of string theoretic duality symmetries the pure cosine potentials of
natural inflation are replaced by modular functions. This leads 
to ``wiggles'' in the inflationary potential
that modify the predictions with respect to CMB-observations.
In particular, the scalar power spectrum deviates from the standard power law form.
As a by-product one can show that trans-Planckian excursions
of the aligned effective axion are compatible with the weak gravity
conjecture.
\end{abstract}
\maketitle

\section{Introduction}

Natural (axionic) inflation~\cite{Freese:1990rb} is one of the best-motivated scenarii
to describe the inflationary expansion of the early universe. The 
flatness of the potential is guaranteed by a shift symmetry that is
perturbatively exact, only broken by non-perturbative (instantonic)
effects. The mechanism can accommodate sizeable primordial tensor modes
that require trans-Planckian excursions of the inflaton field. While
in the simplest form of axionic inflation these large-field excursions
are problematic, a satisfactory solution can be found through
a helical motion of the inflaton as suggested in the schemes of
axion alignment~\cite{Kim:2004rp} or axion monodromy~\cite{Silverstein:2008sg}.

Axions are abundant in string theory constructions that could provide 
a consistent ultra-violet completion of natural inflation. The 
embedding in string theory will have some rather model-independent
implications for the inflationary potential. The non-perturbative
effects responsible for the breakdown of the shift symmetry typically
come from 
instantons or gaugino condensates that in many 
cases can be described by modular functions~\cite{Dixon:1986qv,Hamidi:1986vh,Dixon:1990pc,Blumenhagen:2009qh,
Grimm:2007xm,Grimm:2009ef,Arends:2014qca}. Instead of a pure
cosine-potential we thus expect a more complicated picture including
higher harmonics as subleading instanton effects. This leads to
``wiggles'' in the potential that modify the predictions of the simplest
scheme. In case of a single axion this can be shown to lead to an
upper limit on the axion decay constant~\cite{Banks:2003sx} (as a result of the duality symmetries of string theory).

In the present paper we analyze the framework of aligned natural
inflation~\cite{Kim:2004rp} including modulated potential and higher harmonics.
We show that (in contrast to the single field case) a certain 
amount of trans-Planckian excursion is allowed. This modulated
version of aligned axionic inflation induces corrections to
CMB-observables (in particular the scalar spectral index) which
resolve the (mild) tension of natural inflation with the latest
Planck~\cite{Ade:2015lrj} and BICEP2/Keck Array~\cite{Array:2015xqh} data. As a by-product one can show that modulated
natural inflation (including subleading higher harmonics),
automatically satisfies the mild version of the weak gravity
conjecture~\cite{ArkaniHamed:2006dz,Cheung:2014vva}. This version is expected to hold in string theory for the
aligned axion case even in the case of an effective trans-Planckian decay
constant.

\section{Instantons and Modular Functions}

In supergravity, the axion $\varphi$ is part of a complex field $T=\chi+ i \varphi$, where $\chi$ denotes the saxion. At the perturbative level, axions possess a continuous shift symmetry and hence correspond to flat directions in field space. Non-perturbative (instanton) terms generate a periodic potential for the axions, which in the simplest case takes the form 
\begin{equation}
V=\Lambda^4 \left(1-\cos{\left[\frac{\varphi}{f}\right]}\right)\,,
\end{equation}
where $f$ denotes the axion decay constant.

In many instances, this just represents the leading potential. Indeed, the non-perturbative superpotential often contains a series of higher harmonics. This is well known for toroidal string compactifications where these contributions can be calculated from first principles~\cite{Dixon:1986qv,Hamidi:1986vh,Dixon:1990pc,Blumenhagen:2009qh}. Also in more general string theory setups, higher harmonics populate the non-perturbative superpotential~\cite{Grimm:2007xm,Grimm:2009ef,Arends:2014qca}. The occurrence of these subleading instanton effects in the form of $\eta$- and $\vartheta$-functions has already been considered for inflation model building~\cite{Grimm:2007hs,Ruehle:2015afa,Abe:2014xja,Higaki:2015kta}.
As an illustrative example, we consider couplings in toroidal string compactifications.
Couplings $y$ between chiral matter fields $\phi_\alpha,\dots,\phi_\beta$ 
are generated by 
non-perturbative effects
\begin{equation}\label{eq:orbifoldyukawa}
W\supset y\:\phi_\alpha\cdots\phi_\beta \propto \prod_{i=1}^3 \eta(T_i)^{2n_i}\times\phi_\alpha\cdots\phi_\beta\,,
\end{equation}
where the $n_i$ are determined by 
localization properties of the fields $\phi_\alpha,\dots,\phi_\beta$ (see e.g.~\cite{Lauer:1989ax,Lauer:1990tm,Ibanez:1992hc,Dundee:2010sb} for the case of heterotic orbifolds), and $\eta(T)$ denotes the Dedekind $\eta$-function
\begin{equation}
\eta(T)= e^{-\pi T/12}\times \prod_{k=1}^{\infty} \left( 1 - e^{-2k \pi T}\right)\,.
\end{equation}
After the matter fields eventually receive a vacuum expectation value,
a periodic potential for the axionic components of the $T_i$ arises. The approximation $\eta(T)\simeq e^{-\pi T/12}$ leads to the standard cosine potential. But as we shall discuss in more detail later, the subleading instantons contained in $\eta$ induce ``wiggles'' on the potential.

\section{Axion Potentials}\label{sec:axpotential}

Let us now turn in more detail to the axion potential including the higher harmonics. We will consider a class of models which have a supersymmetric ground state. They contain an additional chiral superfield $\psi$ known as the stabilizer. The superpotential reads
\begin{equation}
W = \psi\,\Big(A\;\eta(T)^{2n} - B\Big)\;,
\end{equation}
where $A$ and $B$ are, in general, functions of chiral fields. We set the latter to their vacuum expectation values and treat $A,B$ as effective constants. This superpotential is motivated from heterotic orbifolds, where $T$ is identified with a K\"ahler modulus~\cite{Ruehle:2015afa}. 
The corresponding K\"ahler potential 
\begin{equation}\label{eq:kahlerpot}
K = -\log(T+\bar{T})+|\psi|^2
\end{equation}
is shift-symmetric.\footnote{The K\"ahler metric of $\psi$ may in general depend on $T$. As the dependence does not affect our results we have neglected it here.} The model has a supersymmetric Minkowski minimum at $T=\eta^{-1}[(B/A)^{1/2n}]\equiv T_0$. We set the saxion to the minimum and $\psi=0$. The potential for the canonically normalized axion $\varphi=\text{Im}(T)/(\sqrt{2}T_0)$ then reads
\begin{equation}
\begin{split}
V &=  \Lambda^4\, e^{-S_1}\left(1- \cos{\left[\frac{\varphi}{f}\right] }\right)\\
&\quad - 2n\,\Lambda^4 \,e^{-S_2}\left(\cos{\left[\frac{\varphi}{f_\text{mod}}\right] }- \cos{\left[\frac{\varphi}{f}+\frac{\varphi}{f_\text{mod}}\right] }\right)
\\
&\quad+\dots\,,
\end{split}
\end{equation}
with $\Lambda^4= A B/T_0$. The decay constants are given as
\begin{equation}
f=\frac{3\sqrt{2}}{n\pi T_0}\,, \qquad f_\text{mod} =\frac{1}{2\sqrt{2}\pi T_0}= \frac{n}{12}f\,.
\end{equation}
The instanton actions
\begin{equation}\label{eq:instantonaction}
S_1=\frac{n\pi }{6} T_0\,,\qquad
S_2=\left(\frac{n\pi }{6}+2\pi\right)T_0\,
\end{equation}
correspond to the leading and next-to-leading term in the expansion of the Dedekind $\eta$-function. The dots include further, more subleading, terms from the $\eta$-function expansion.

We thus find that higher harmonics in $\eta$ show up with growing frequency and (exponentially) decreasing amplitude in the potential. Another interesting observation is that the amplitude of the wiggles on the potential is controlled by the leading potential. We may write
\begin{equation}
V = \Lambda^4\, e^{-S_1}\left(1- \cos{\left[\frac{\varphi}{f}\right] }\right)\times F(\varphi)\,,
\end{equation}
where we defined
\begin{equation}
\begin{split}
F(\varphi) &= 1 - \delta\: \frac{\sin\left[\tfrac{\varphi}{f_\text{mod}}+\tfrac{\varphi}{2f}\right]}{\sin\left[\tfrac{\varphi}{2f}\right]}\\
&\simeq 1-\delta\cos\left[\frac{\varphi}{f_\text{mod}}\right]\,.
\end{split}
\end{equation}
with $\delta = 2 n \,e^{-2\pi T_0}$. The last equality holds in the vicinity of the maximum of the potential which is mainly relevant in this work.\footnote{We will be mainly concerned with modulations at field values $\varphi_*$, where the scales observed in the CMB cross the horizon. In natural inflation, the approximation holds at $\varphi_*$ if we assume $f<10$.}

The single-axion model is not suitable for inflation as it requires a 
trans-Planckian axion decay constant. With growing $f$ the wiggles on the potential caused by the higher harmonics become more pronounced. For $f\gtrsim 1$ they are no longer suppressed and spoil the potential.\footnote{Throughout this paper we work in units where $M_\text{P}=1$.} This is in agreement with the general arguments presented in~\cite{Banks:2003sx}. 

\section{Aligned Natural Inflation with Modulations}

\subsection{Alignment Mechanism}\label{sec:alignmechanism}
Let us now generalize the above considerations to the case of two axions and two stabilizers. We generalize the superpotential from~\cite{Kappl:2015pxa} by the inclusion of higher harmonics
\begin{equation}
\begin{split}
\label{eq:2fields}
W &= \psi_1\: \Big(A_1\, \eta(T_1)^{2n_1}\eta(T_2)^{2n_2}-B_1\Big) \\
&\quad+\psi_2\: \Big(A_2\, \eta(T_1)^{2m_1}\eta(T_2)^{2m_2}-B_2\Big)\;,
\end{split}
\end{equation}
with two K\"ahler moduli $T_i$ and two stabilizers $\psi_i$. The K\"ahler potential contains two copies of~\eqref{eq:kahlerpot}. The supersymmetric minimum is located at $\psi_i=0$ and $T_i=T_{i,0}$ with $T_{i,0}$ such that in~\eqref{eq:2fields} the terms in brackets vanish.
For the moment we neglect all subleading terms in the expansion of the $\eta$-function which effectively then results in the model discussed in~\cite{Kappl:2015pxa}. Further, we assume that the saxions stay at their minima. This allows us to define the canonically normalized axions as $\varphi_i=\text{Im}(T_i)/(\sqrt{2}T_{i,0})$. The axion potential reads
\begin{equation}
\begin{split}
V &= \Lambda_a^4\, e^{-S_a}\left(1- \cos{\left[\frac{\varphi_1}{f_1}+\frac{\varphi_2}{f_2}\right] }\right)\\
&\quad+\Lambda_b^4\, e^{-S_b}\left(1- \cos{\left[\frac{\varphi_1}{g_1}+\frac{\varphi_2}{g_2}\right] }\right)\;,
\end{split}
\end{equation}
where we have defined  $\Lambda_a^4= A_1\,B_1/(2T_{1,0}T_{2,0})$ and $\Lambda_b^4= A_2\,B_2/(2T_{1,0}T_{2,0})$ . The instanton actions read
\begin{align}
S_a&=\frac{(n_1T_{1,0}+n_2T_{2,0})\,\pi }{6} \,,\label{eq:insac}\\
S_b&=\frac{(m_1T_{1,0}+m_2T_{2,0})\,\pi }{6} \,.
\end{align}
The axion decay constants are given as
\begin{align}
f_i= \frac{3\sqrt{2}}{n_i\pi T_{i,0}}\,,\quad
g_i= \frac{3\sqrt{2}}{m_i\pi T_{i,0}}\,.
\end{align}
If we assume for simplicity that $\Lambda_b^4\, e^{-S_b}\gg \Lambda_a^4\, e^{-S_a}$, there appears a heavy ($\tilde{\varphi}\propto \tfrac{\varphi_1}{g_1}+\tfrac{\varphi_2}{g_2}$) and a light ($\varphi\propto \tfrac{\varphi_1}{g_2}-\tfrac{\varphi_2}{g_1}$) linear combination of axions. After integrating out the heavy combination, the potential becomes
\begin{equation}
V= \Lambda_a^4\, e^{-S_a}\left(1- \cos{\left[\frac{\varphi}{f}\right] }\right)
\end{equation}
with the effective axion decay constant
\begin{equation}
\begin{split}
\label{eq:effdecay}
f&= \sqrt{g_1^2+g_2^2}\;\,\frac{f_1 f_2}{g_1 f_2-g_2 f_1}\\
&=\frac{3\sqrt{2}}{\pi(n_1 m_2 - m_1 n_2)}\;\sqrt{\frac{m_1^2}{T_{2,0}^2}+\frac{m_2^2}{T_{1,0}^2}}\,.
\end{split}
\end{equation}
Even with all individual decay constants being sub-Planckian, a trans-Planckian $f$ can be realized. This is achieved if the alignment condition~\cite{Kim:2004rp} (see~\cite{Choi:2014rja} for the generalization to many axions)
\begin{equation}
\frac{f_1}{f_2}\simeq \frac{g_1}{g_2}\quad \Longleftrightarrow\quad\frac{n_1}{n_2}\simeq \frac{m_1}{m_2}\;.
\end{equation}
is met. The effective decay constant is altered if the saxions $\text{Re}(T_i)$ are displaced during inflation. At the supersymmetric minimum the light axion and saxion linear combinations are mass-degenerate. However, as we discussed in detail in~\cite{Kappl:2015pxa}, the saxion receives a large Hubble mass term during inflation which decouples it from the inflationary dynamics. The saxion displacement during inflation is negligible if $m_1 T_{1,0} \sim m_2 T_{2,0}$. We will concentrate on this case such that~\eqref{eq:effdecay} approximately holds (see~\cite{Kappl:2015pxa}).

\subsection{The Weak Gravity Conjecture}

Some concerns on the axion alignment mechanism have been raised in~\cite{Rudelius:2014wla,delaFuente:2014aca,Rudelius:2015xta,Montero:2015ofa,Brown:2015iha,
Brown:2015lia,Junghans:2015hba,Heidenreich:2015wga,Heidenreich:2015nta} on the basis of the weak gravity conjecture. The conjecture was originally introduced in~\cite{ArkaniHamed:2006dz} to constrain U(1) gauge interactions and generalized in~\cite{Cheung:2014vva}. It was noted that a U(1) theory should contain a particle with charge-to-mass ratio $q/m > 1$ such that extremal black holes can decay. Otherwise, there would be an infinite number of stable gravitational bound states which might cause problems with the covariant entropy bound~\cite{Banks:2006mm}. 
The so-called mild version of the weak gravity conjecture states that a particle with the charge-to-mass ratio $q/m > 1$ must exist, whereas the strong version further insists that it should be the lightest charged particle. The evidence for the strong version is rather weak and we therefore consider in the following only the mild version of the weak gravity conjecture.
In string theory, gauge fields can be linked to axions via dualities~\cite{Brown:2015iha}. According to the dictionary, the inverse axion decay constant plays the role of the charge and the instanton action plays the role of the mass. 

The weak gravity conjecture can be generalized to multi-axion systems. Given a Lagrangian with $\alpha$ axions and $i$ instanton terms
\begin{equation}
\mathcal{L}= \frac{1}{2} \partial_\mu \varphi_\alpha\partial^\mu\varphi_\alpha - \Lambda_i^4\, e^{-\mathcal{S}_i}\left(1- \cos{\left[c_{i,\alpha}\,\varphi_\alpha\right] }\right)
\end{equation}
one needs to determine the convex hull spanned by the vectors $\pm (c_{i,\alpha}/\mathcal{S}_i)$. 
The weak gravity conjecture is satisfied if the convex hull contains a ball of radius $\mathcal{O}(1)$ called the `unit ball'. The exact radius depends on the type of axion under consideration, some examples are discussed in~\cite{Brown:2015iha}.

\begin{figure}[t]
\centering{\includegraphics[width=6cm]{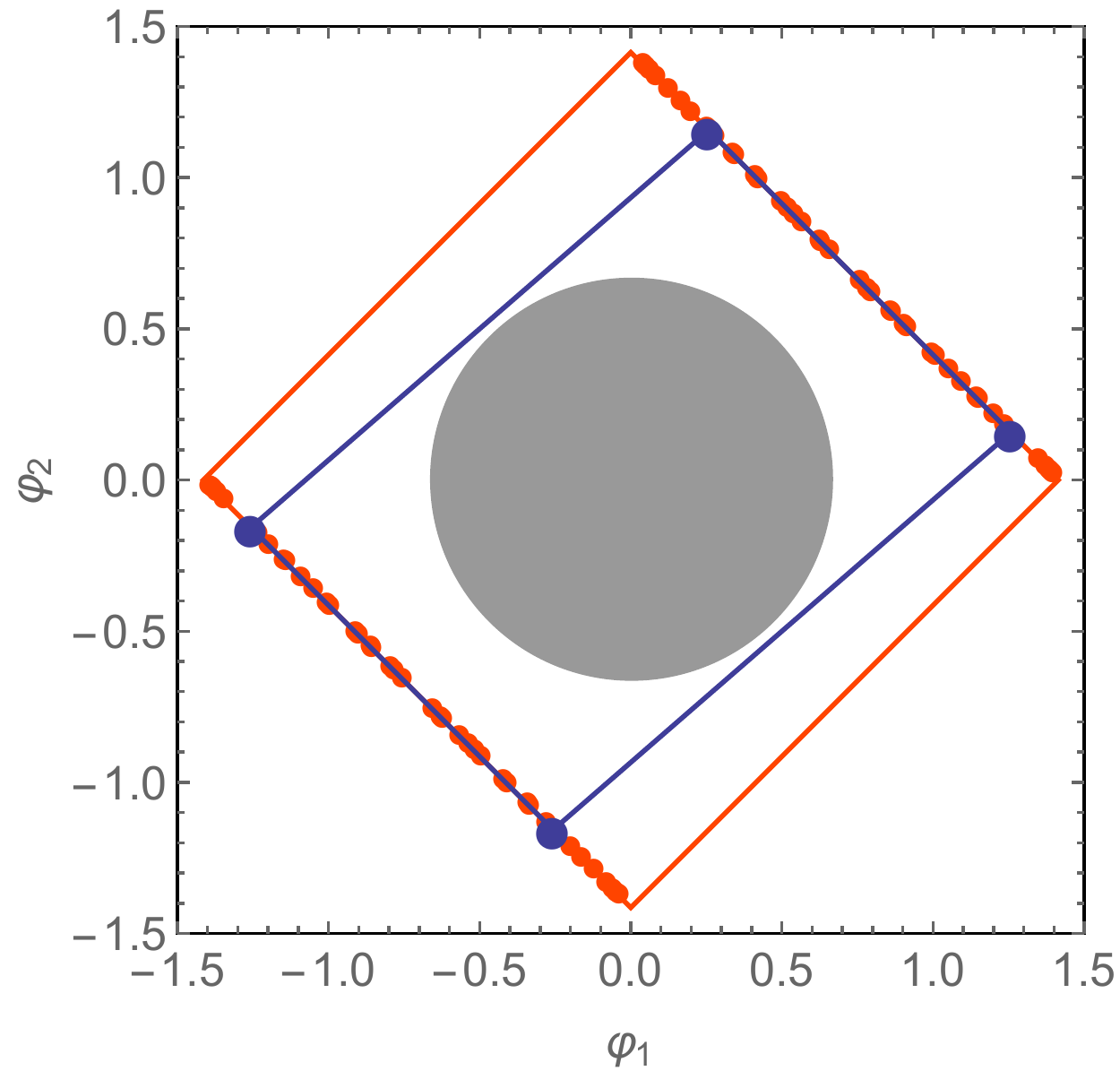}}\\[4mm]
\centering{$\qquad\quad$\includegraphics[width=5cm]{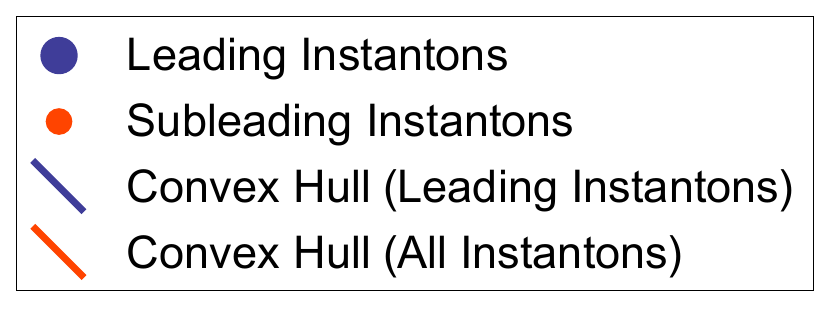}}\\[5mm]
\centering{\includegraphics[width=6cm]{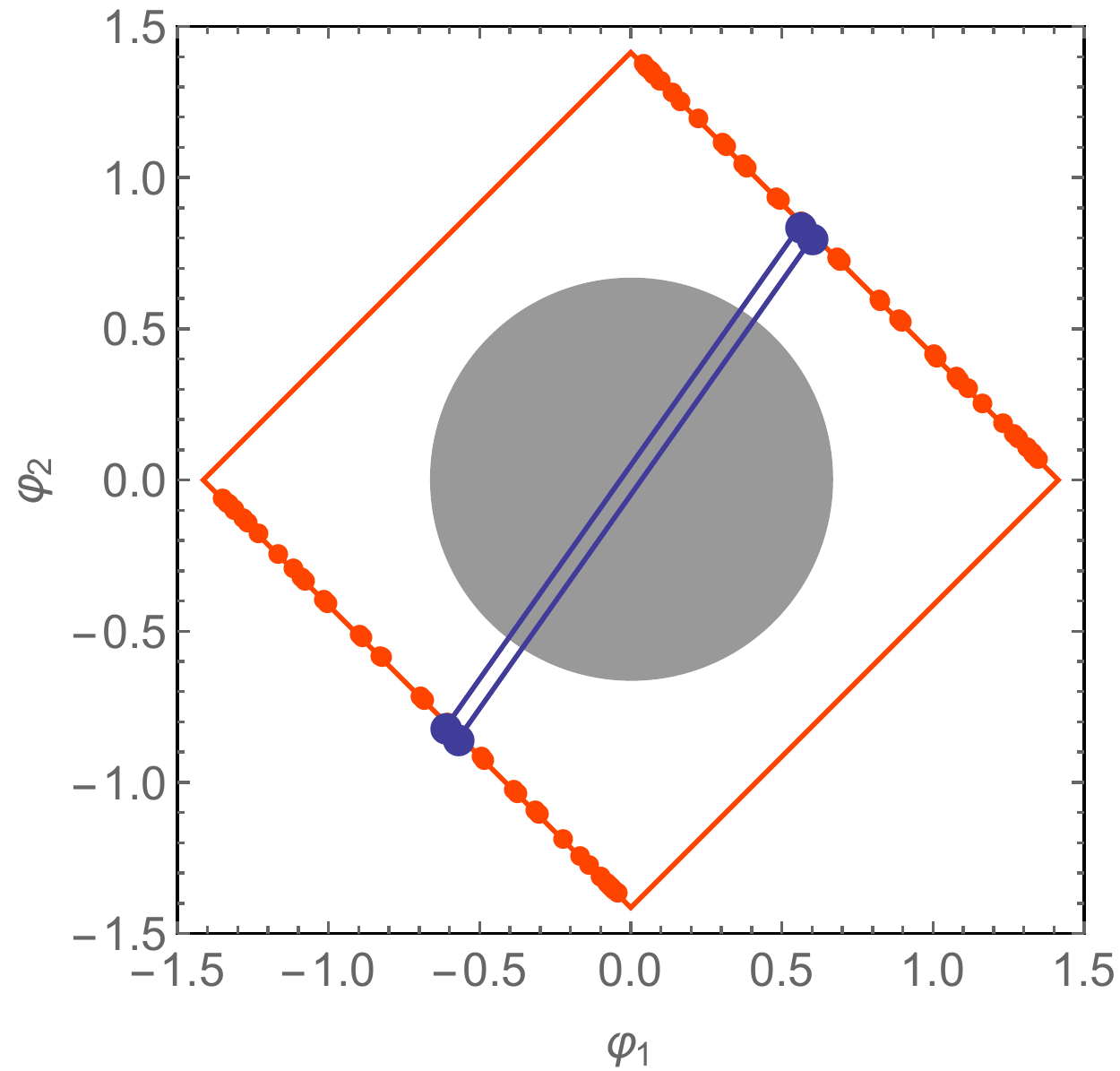}}
\caption{Convex hull for the model~\eqref{eq:2fields} including the leading and all instantons. To avoid clutter, subleading instantons are only shown up to order 10 in the expansion of the $\eta$-function. The upper panel refers to a case of decay constant misalignment, the lower panel to alignment. In the alignment case, the convex hull condition is clearly violated if one considers only the leading instantons. However, by taking into account all subleading instantons, the convex hull dramatically increases. The weak gravity conjecture can thus be satisfied even in the alignment case. For illustrative purposes we have included a representative `unit ball'. See~\cite{Brown:2015iha} for discussion of the size of the `unit ball'.}
\label{fig:hull}
\end{figure}

The simplest aligned axion inflation model with just two instantons (two cosine terms) violates the weak gravity conjecture. In order to illustrate this, let us turn to the model discussed in the previous section. The charge vectors for the leading instantons are given as
\begin{subequations}\label{eq:chargevector}
\begin{align}
v_1 &= \frac{\sqrt{2}}{n_1 T_{1,0}+n_2 T_{2,0}} \;\begin{pmatrix}n_1 T_{1,0}\\[1mm]n_2 T_{2,0}\end{pmatrix}\\
v_2 &= \frac{\sqrt{2}}{m_1 T_{1,0}+m_2 T_{2,0}} \;\begin{pmatrix}m_1 T_{1,0}\\[1mm]m_2 T_{2,0}\end{pmatrix}
\end{align}
\end{subequations}
In figure~\ref{fig:hull}, we depict the convex hull spanned by these vectors. We have chosen two representative examples with alignment ($n_1=3$, $n_2=4$, $m_1=2$, $m_2=3$) and misalignment  ($n_1=8$, $n_2=1$, $m_1=2$, $m_2=9$). In the alignment case, the charge vectors are almost parallel. The convex hull condition seems to be violated.

However, there arise further instantons which are subdominant in the potential. Still, they may significantly contribute to the convex hull~\cite{Bachlechner:2015qja,Hebecker:2015rya}. By expanding the $\eta$-function in~\eqref{eq:2fields}, one verifies that the charge vectors of the subleading instantons are obtained from $v_1$ in~\eqref{eq:chargevector} by substituting 
\begin{equation}
n_1\rightarrow n_1 + 12 \,k_1\,,\quad n_2\rightarrow n_2 + 12 \,k_2\,,\quad k_{1,2} \in \mathbb{N}\,,
\end{equation}
and analogously from $v_2$. In figure~\ref{fig:hull}, we also show the convex hull including all subleading instantons. 

It can be seen that in both cases -- alignment or misalignment -- the same convex hull is obtained. While we refrain from determining the model-dependent radius of the `unit ball', the convex hull condition is assumed to be fulfilled in the case of decay constant misalignment.  Hence, it must also be satisfied in the alignment case by including the subleading instantons. The presented model is thus consistent with the weak gravity conjecture. The $\eta$-function automatically implements the mechanism outlined in~\cite{Bachlechner:2015qja,Hebecker:2015rya}. We conclude that through the unavoidable occurrence of subleading instantons, string theory satisfies the weak gravity conjecture automatically.

\subsection{Modulated Natural Inflation}
The higher harmonics can reconcile the alignment mechanism with the weak gravity conjecture. But at the same time, they must be sufficiently suppressed in order not to spoil the flatness of the inflaton potential. In the single-axion case (section~\ref{sec:axpotential}), the higher harmonics are controlled only for $f\ll 1$. This raises the question if $f>1$ can be achieved in the two-field case.

\begin{figure}[htp]
\begin{center}
  \includegraphics[height=8cm]{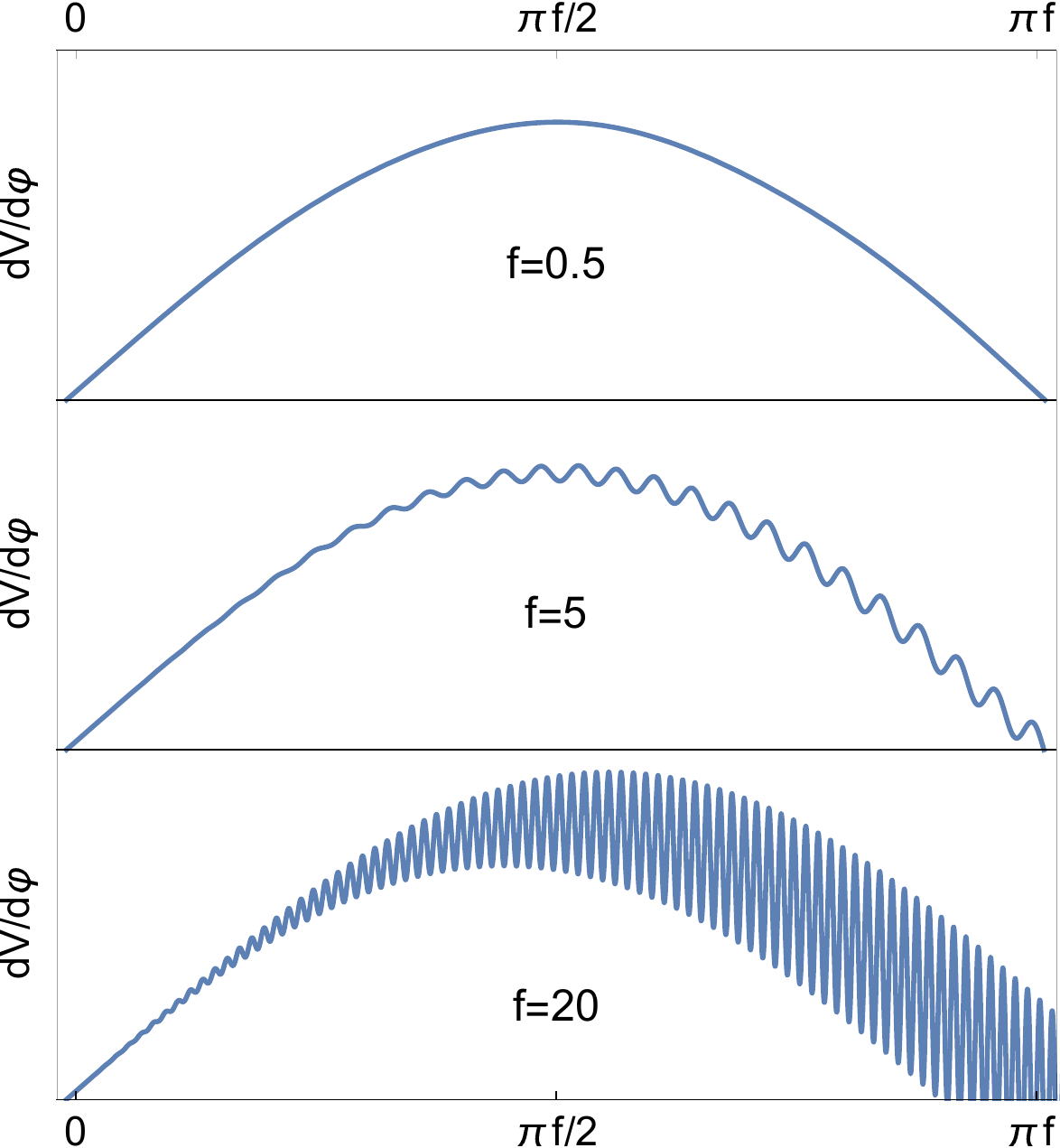}
\end{center}
\caption{Slope of the axion potential for three exemplary parameter choices. The alignment increases from top to bottom. For too strong alignment the potential (and its derivatives) is very ``wiggly'' and unsuitable for inflation.}
\label{fig:axpot}
\end{figure}

We estimate the leading correction to the axion potential arising from the second term in the expansion of the $\eta$-function. As in section~\ref{sec:alignmechanism}, we assume $\Lambda_b^4\, e^{-S_b}\gg \Lambda_a^4\, e^{-S_a}$. We consider the case\footnote{In the opposite case $T_2 > T_1$ one just needs to exchange the indices $1$ and $2$ in the following discussion.} $T_1 >  T_2$ such that the dominant modulation arises by expanding $\eta(T_2)^{2 n_2}$. Taking the leading correction into account, the potential for the light axion becomes
\begin{equation}\label{eq:effaxpot}
V \simeq \Lambda^4\left(1- \cos{\left[\frac{\varphi}{f}\right] }\right) \left(1-\delta\cos\left[\frac{\varphi}{f_\text{mod}}\right]\right)\,,
\end{equation}
with $f$ as defined in~\eqref{eq:effdecay} and
\begin{equation}
\Lambda^4 \simeq \Lambda_a^4\, e^{-S_a} = \frac{A_1\,B_1}{2T_{1,0}T_{2,0}}
\,e^{-(n_1T_{1,0}+n_2T_{2,0})\,\pi /6}\,.
\end{equation}
The modulation frequency can be estimated as
\begin{equation}
  f_\text{mod} = \frac{\sqrt{m_1^2+m_2^2}}{m_1}\; \frac{1}{2\sqrt{2}\pi T_{2,0}}\,,
\end{equation}
and the relative amplitude is
\begin{equation}
\delta = 2 n_2 \,e^{-2\pi T_{2,0}}\,.
\end{equation}
Notice that for decay constant alignment, only the period $f$ of the leading potential is enhanced, but not the period of the wiggles $f_{\text{mod}}$. This is illustrated in figure~\ref{fig:axpot} where we show the slope of the axion potential for three representative benchmark examples. Clearly the higher harmonics limit the achievable enhancement of the axion decay constant.
In order to show that a trans-Planckian $f$ can still be obtained, we now turn to a numerical example.

\section{Numerical Example}
We consider the model~\eqref{eq:2fields} with the parameters of table~\ref{tab:benchmark}.
\begin{table}[htp]
\begin{center}
\begin{tabular}{cccccccccc}
\hline
$n_1$ & $n_2$ & $m_1$ & $m_2$ &$A_1$ & $A_2$ & $B_1$ & $B_2$ & $T_{1,0}$ &$T_{2,0}$  \\
\hline
$4$ & $5$ & $3$ & $4$ &$0.18$ & $0.117$ & $3.6\cdot 10^{-5}$ & $1.6\cdot 10^{-4}$ & $2.0$ & $1.65$\\
\hline
\end{tabular}
\end{center}
\caption{Parameter choice leading to successful inflation.}
\label{tab:benchmark}
\end{table}
This is a realistic parameter choice in heterotic orbifolds (see e.g.\cite{Ruehle:2015afa})\footnote{It was shown~\cite{Palti:2015xra} that the embedding in type IIA compactifications seems to be difficult.}. Due to the alignment, the effective decay constant for the lightest axion is enhanced.
\begin{figure}[htp]
\centering{\includegraphics[width=7cm]{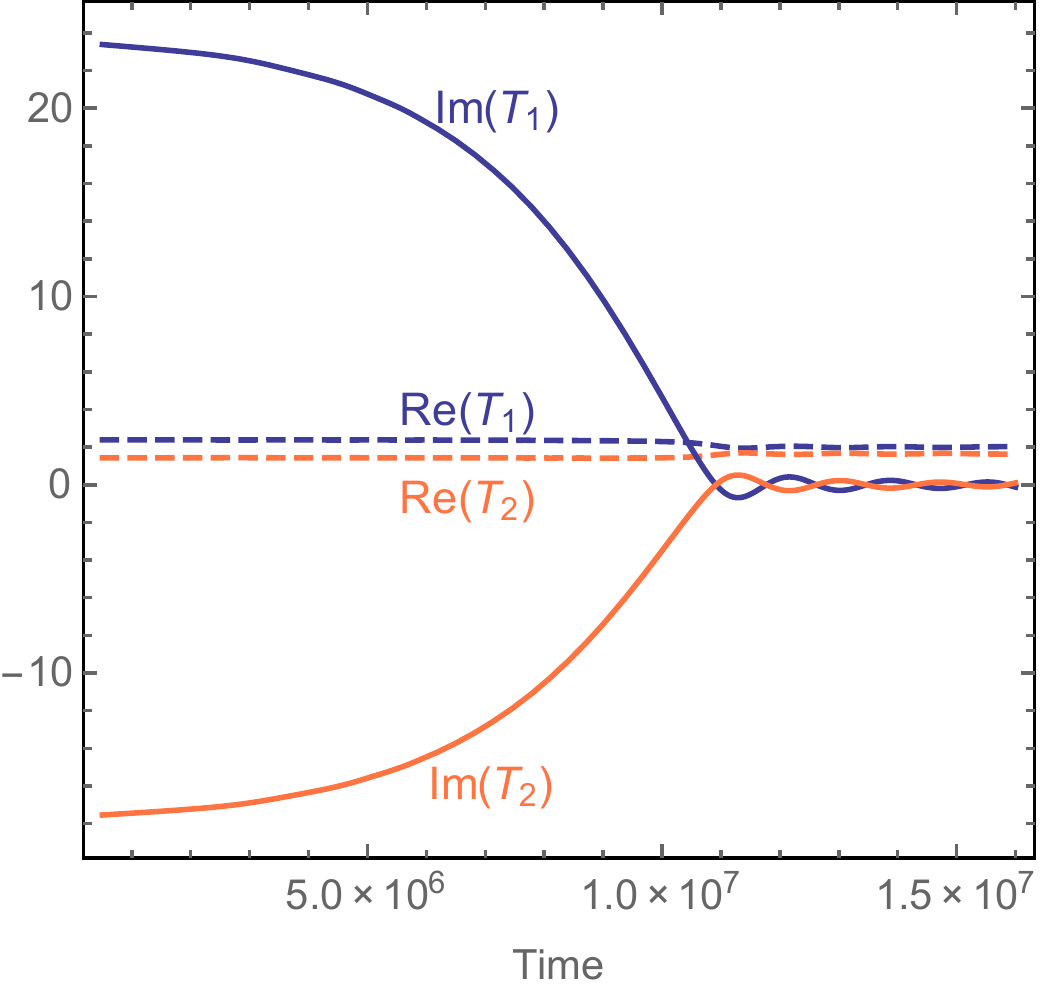}}
\caption{Trajectories of the axions and saxions during inflation.}
\label{fig:trajectories}
\end{figure}
We have solved the equations of motion for this multi-field system~\cite{Sasaki:1995aw}
\begin{equation}
\ddot{\phi}^\alpha+\Gamma^{\alpha}_{\beta\gamma}\dot{\phi}^{\beta}\dot{\phi}^{\gamma}+3H\dot{\phi}^{\alpha}+G^{\alpha\beta}\frac{\partial V}{\partial\phi^{\beta}}=0\,. 
\end{equation}
Here the fields \(\phi^{\alpha}\) label the real and imaginary parts of $T_{1,2}$. The field space metric \(G_{\alpha\beta}\) can be determined from the K\"ahler metric
and 
\(\Gamma^{\alpha}_{\beta\gamma}\) is the usual Christoffel symbol with respect to the field metric \(G_{\alpha\beta}\) and its inverse \(G^{\alpha\beta}\).
In figure~\ref{fig:trajectories} we depict the trajectories of the real and imaginary parts of $T_{1,2}$ during inflation. The inflaton can be identified with a linear combination of axions, while the remaining fields reside close to their minima. Indeed, we find that this model effectively reduces to a single field inflation model. The potential is well approximated by~\eqref{eq:effaxpot} with the parameters given in table~\ref{tab:out}.
\begin{table}[htp]
\begin{center}
\begin{tabular}{cccc}
\hline
$\Lambda$ & $f$ & $f_\text{mod}$ & $\delta$  \\
\hline
$3.6\cdot 10^{-3}$ & $3.64$ & $0.10$ & $3.3\cdot 10^{-4}$ \\
\hline
\end{tabular}
\end{center}
\caption{Parameters of the inflation potential~\eqref{eq:effaxpot} obtained in the benchmark model.}
\label{tab:out}
\end{table}
We have obtained the parameters by fitting the functional form~\eqref{eq:effaxpot} to the potential along the inflationary valley. They are, however, in very good agreement with the analytic estimates presented in the previous section.

\begin{figure}[htp]
\centering{\includegraphics[width=7cm]{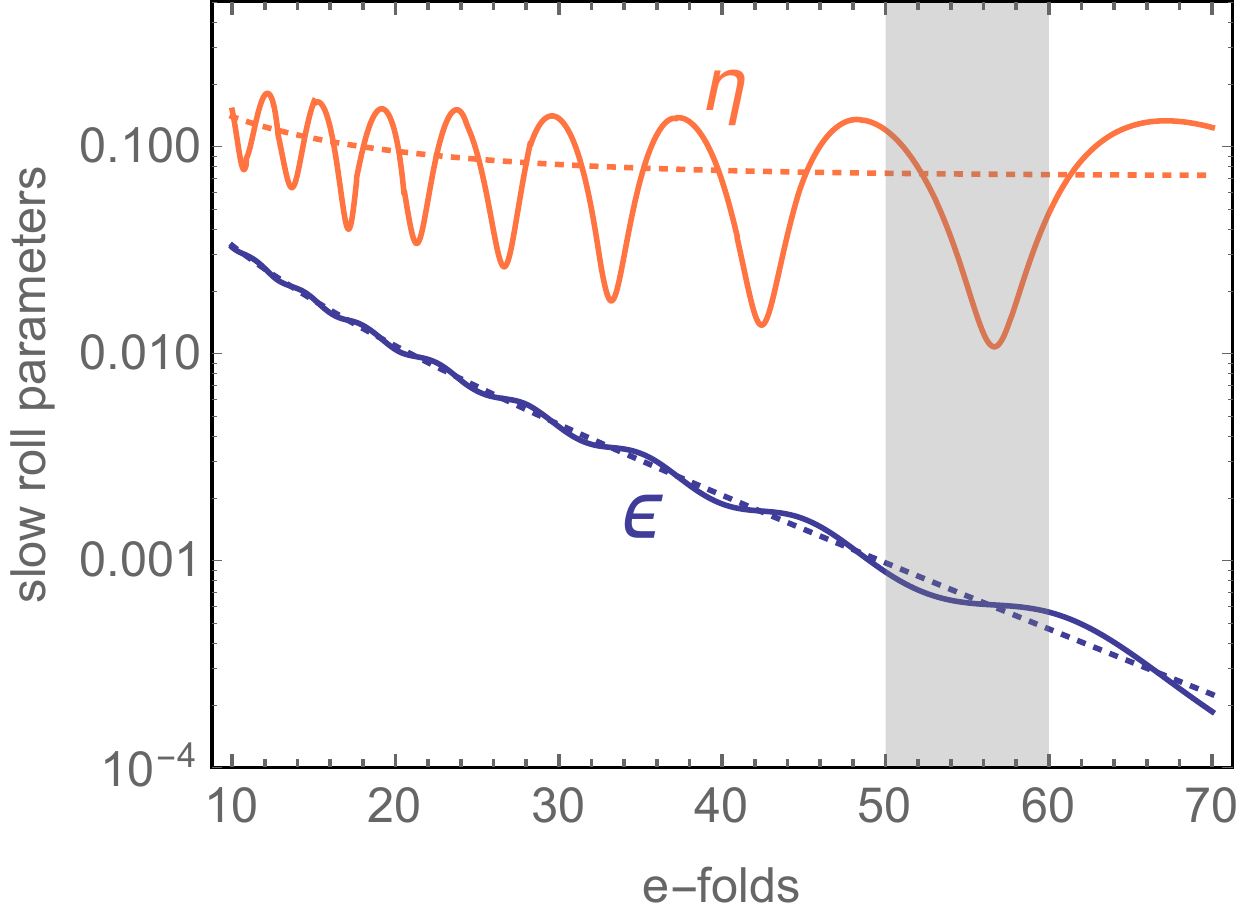}}
\caption{Slow roll parameters $\epsilon$ (blue) and $\eta$ (orange) as a function of e-folds. For the dotted lines we neglected subleading instantons (which cause the wiggles on the potential).}
\label{fig:slowroll}
\end{figure}

The small wiggles on the potential can hardly be seen by eye, but they have important implications. In figure~\ref{fig:slowroll} we depict the slow roll parameters~\cite{Baumann}
\begin{equation}
\epsilon=-\frac{\dot{H}}{H^2}\,, \qquad \eta = \frac{\dot{\epsilon}}{\epsilon H}\,.
\end{equation}
as a function of the e-folds.
The slow roll parameters oscillate due to the wiggles on the potential. While $\epsilon$ is mildly affected, there is a stronger effect on $\eta$. This can be understood as $\eta$ is related to the second derivative of the potential. The modulations affect the curvature of the potential much stronger than the slope.

In order to compare with observation we have calculated the scalar and tensor power spectra $\mathcal{P}_{\mathcal{R}}(k)$ and $\mathcal{P}_{t}(k)$ for the benchmark scenario. Due to the modulations on the potential, we did not fully trust the slow-roll approximation. Therefore, we numerically solved the full Mukhanov-Sasaki mode equations. The CMB-observables are then directly evaluated from the power spectra. We solve only for the quantum fluctuations in the direction of the inflaton and neglect possible orthogonal entropy contributions. With this simplification we only have to deal with a single field system for the mode equation, whereas the background equations are still solved for all fields. The Mukhanov-Sasaki mode equation in conformal time \(d\tau = dt/a\) can be written as (see e.g.~\cite{Baumann})
\begin{equation}
\frac{d^2u_k}{d\tau^2}+\left(k^2-\frac{1}{z}\frac{d^2z}{d\tau^2}\right)u_k=0\,,
\end{equation}
with \(z^2=2a^2\epsilon\) and
\begin{align}
\frac{1}{z}\frac{d^2z}{d\tau^2}&=(aH)^2\left(2-\epsilon+\frac{3}{2}\eta-\frac{1}{2}
\epsilon\eta+\frac{1}{4}\eta^2+\eta\kappa\right)\,,\\ 
\kappa&=\frac{\dot{\eta}}{H\eta}\,.
\end{align}
We solved the equation for each mode \(k\) starting a few e-folds before horizon crossing of the given mode until the end of inflation. As initial conditions the standard Bunch-Davies vacuum~\cite{Bunch:1978yq}
\begin{equation}
u_k=\frac{1}{\sqrt{2k}}e^{-ik\tau}
\end{equation}
has been applied. The solutions \(u_k\) define the scalar power spectrum by 
\begin{equation}
\mathcal{P}_{\mathcal{R}}(k)=\frac{k^3}{2\pi^2}\left|\frac{u_k}{z}\right|^2\,.
\end{equation}
The spectral index is then defined as
\begin{equation}
n_s=1+\frac{d\log\mathcal{P}_{\mathcal{R}}(k)}{d\log k}\,.
\end{equation}
The tensor mode equation and its power spectrum are given by
\begin{equation}
\frac{d^2v_k}{d\tau^2}+\left(k^2-\frac{1}{a}\frac{d^2a}{d\tau^2}\right)v_k=0\,,
\qquad
\mathcal{P}_{t}(k)=\frac{k^3}{2\pi^2}\left|\frac{v_k}{a}\right|^2\,.
\end{equation}
We solve this equation like the scalar mode equation implying the same initial conditions. The tensor-to-scalar ratio can then be determined from
\begin{equation}
r=\frac{\mathcal{P}_{t}(k)}{\mathcal{P}_{\mathcal{R}}(k)}\,.
\end{equation}
Similar to the outlined treatment an analogous calculation using directly e-fold time \(N\) would also be possible (see e.g.~\cite{Mortonson:2010er,Price:2014xpa}).

\begin{figure}[t]
\centering{\includegraphics[width=8.5cm]{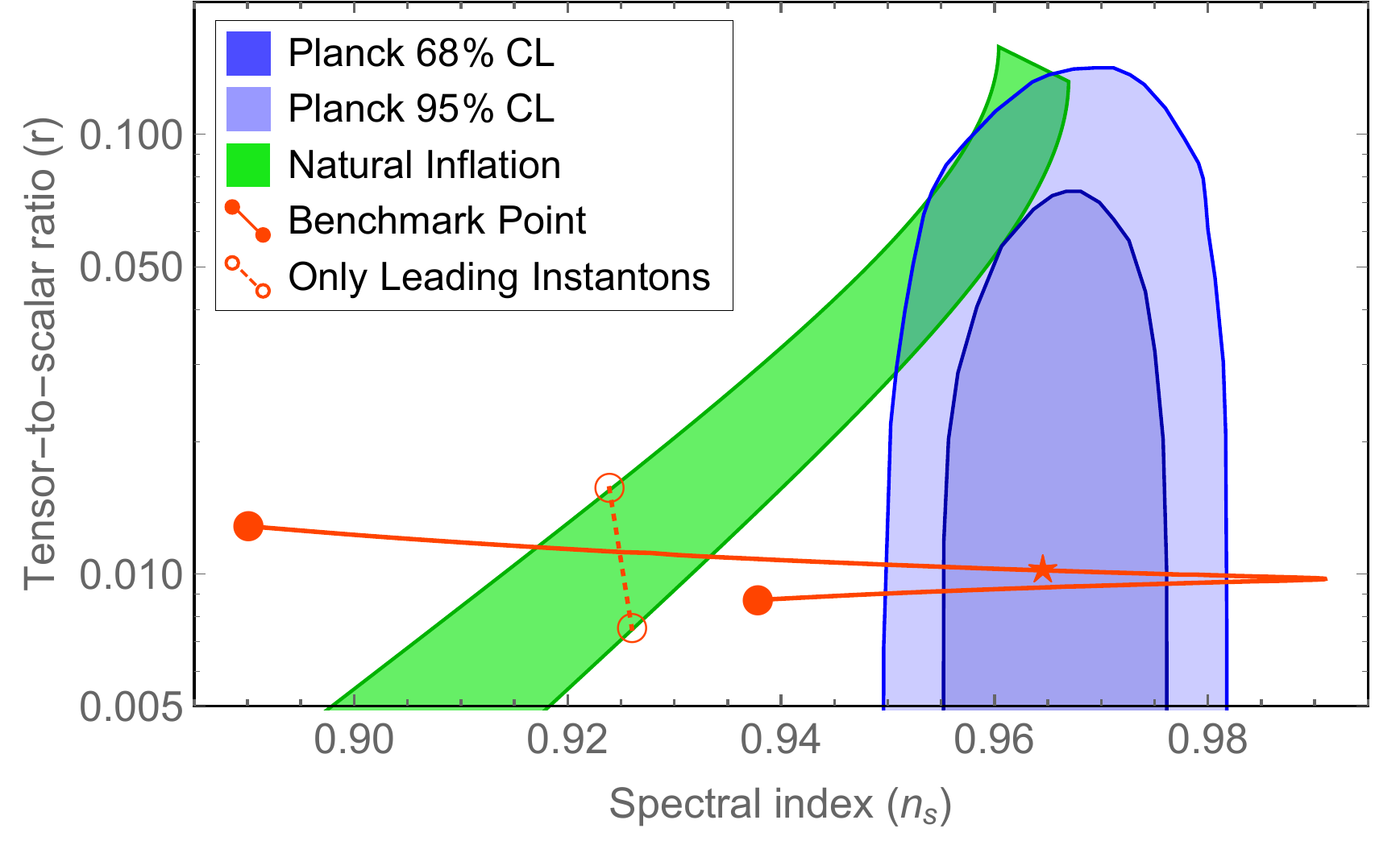}}\\
\centering{\includegraphics[width=8.8cm]{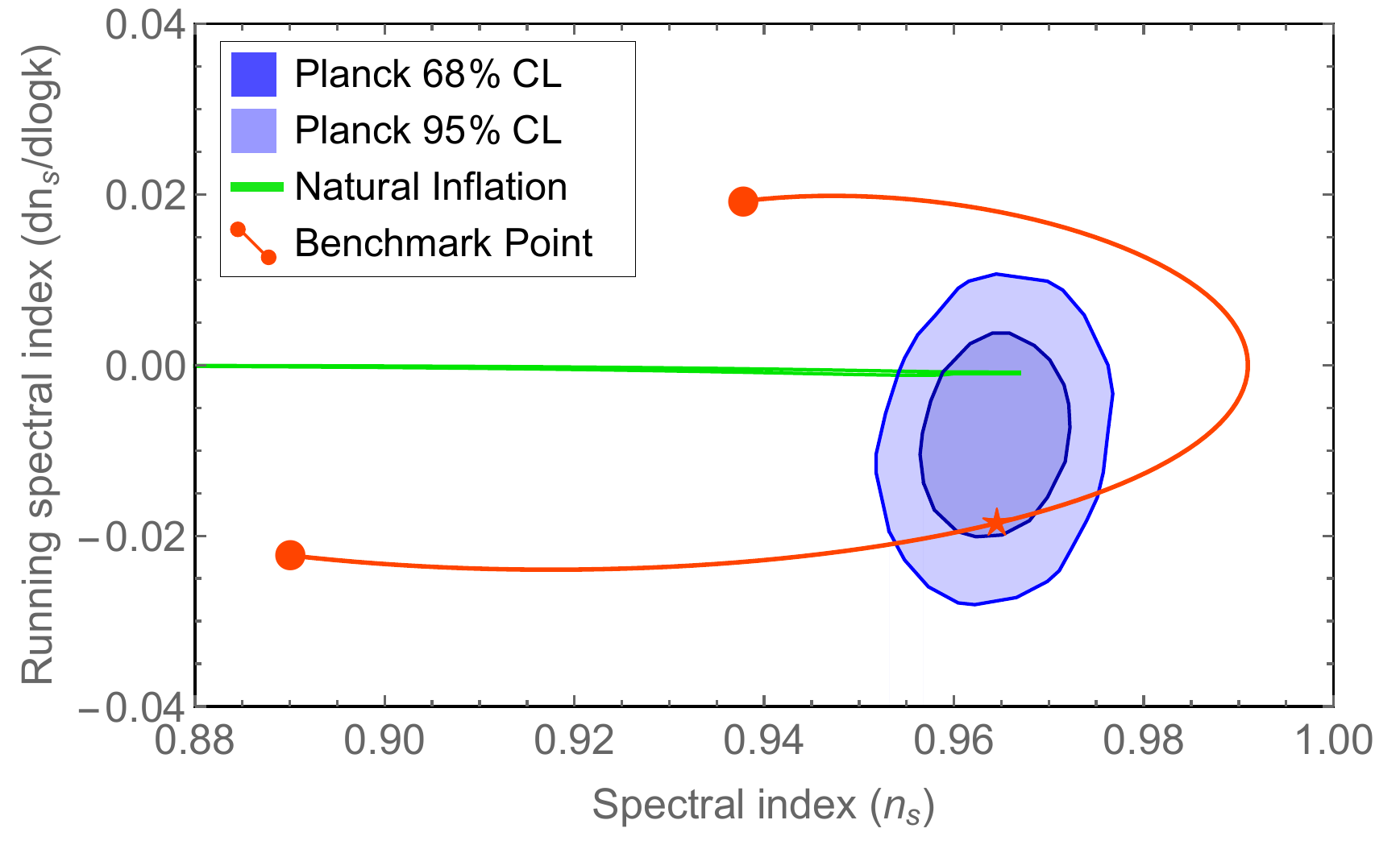}}
\caption{Upper panel: spectral index and tensor-to-scalar ratio for the benchmark scenario assuming $N_*=50-60$ (solid orange line).  The dotted orange line is obtained by neglecting the wiggles on the potential. Also shown is the natural inflation band and the most recent Planck constraints. We depict the Planck constraints in the presence of a running spectral index which are weaker than the constraints without running. Lower panel: spectral index and running of the spectral index for the benchmark scenario (orange line) together with the Planck constraints.  In pure natural inflation the running is negligible (green line).}
\label{fig:cmb}
\end{figure}

In figure~\ref{fig:cmb} we depict the resulting CMB-observables for the benchmark scenario. If we would have neglected the wiggles on the potential, the predictions would lie on the natural inflation band. Natural inflation is in mild tension with the latest Planck data~\cite{Ade:2015lrj}\footnote{See~\cite{Peloso:2015dsa,Achucarro:2015caa} for different ways to relax the tension between natural inflation and observations.}. The modulations on the potential only slightly affect $r$ (which is mainly determined by $\epsilon$), but have huge impact on $n_s$. Due to a substantial running of the spectral index, the prediction on $n_s$ depends strongly on which number of e-folds $N_*$ is matched to the Pivot scale $k_*$. We take $N_* =50-60$ as the range of uncertainty.

\begin{figure*}[htp]
\centering{\includegraphics[width=12cm]{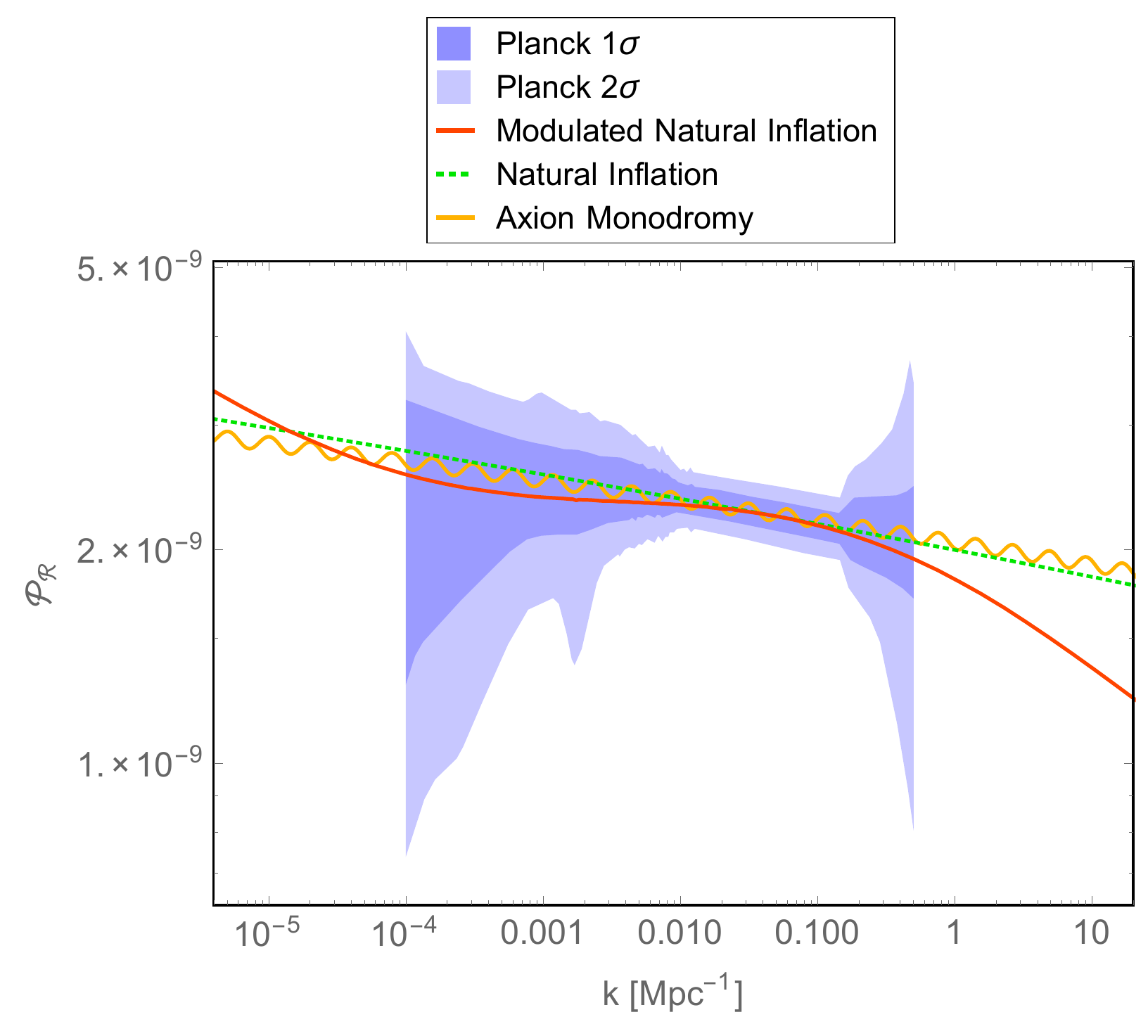}}
\caption{Scalar power spectrum of modulated natural inflation compared to the Planck reconstructed power spectrum. Also shown are representative power spectra of natural inflation and axion monodromy. The concrete form of the power spectrum for axion monodromy is of course model dependent~\cite{Flauger:2009ab}.}
\label{fig:power}
\end{figure*}

Remarkably, the prediction of the benchmark scenario passes through the Planck $68\%$~CL contour. Hence, the modulations on the potential provide a viable possibility to resolve the tension of natural inflation with Planck. While there is a sizeable running of the spectral index, it is shown in the lower panel of figure~\ref{fig:cmb} that this is still consistent with the Planck data.

We should mention, however, that the Planck constraints on running are strictly applicable only in absence of running of the running, i.e.\ for $d^2 n_s/d \log k^2=0$. In our case, the modulations in the potential also induce a non-vanishing $d^2 n_s/d \log k^2$. To check the model more rigorously, we directly compare the predicted scalar power spectrum with the Planck reconstructed power spectrum (from the Bayesian analysis, figure 26 in~\cite{Ade:2015lrj}). For concreteness, we matched the Pivot scale to $N_*=53.5$ indicated by the stars in figure~\ref{fig:cmb}. The result is shown in figure~\ref{fig:power}. Although the predicted power spectrum deviates considerably from the standard power law form, it is fully consistent with Planck. Indeed, the modulations could play a role in explaining the suppression of power at large angular scales (small $k$) favored by the Planck data. For illustration, we also depict two typical power spectra for natural inflation and axion monodromy~\cite{Flauger:2009ab}.

The three power spectra are markedly different: in natural inflation, the spectrum has the standard power law form. In axion monodromy, there arise modulations which, however, come with higher frequency than in modulated natural inflation~\cite{Flauger:2009ab} (see~\cite{Choi} for high frequency modulations in the case of aligned axion inflation).

\section{Conclusion}

Modulated natural inflation appears as a result of embedding axionic
inflation in large classes of string theory. Higher harmonics 
(required by duality symmetries of string theory) cause wiggles in
the leading cosine potential that could strongly influence the
inflationary process. In the single axionic case they prevent
trans-Planckian values of the string decay constant.

In the case of aligned (multi-) axion systems these wiggles still 
allow a certain amount of trans-Planckian excursions of the aligned
effective axion, but the modulations have strong influence on the
CMB-predictions of the scheme. They lead to a substantial running
of the spectral index that alleviates the mild tension of simple
natural inflation with Planck data. Modulated natural inflation
is thus perfectly consistent with all presently known CMB-observations.

Given the substantial running of the spectral index we have performed
(numerical) calculations beyond the slow-roll approximation by
solving the full Mukhanov-Sasaki mode equations. The results for our
benchmark model are shown in figures~\ref{fig:cmb} and~\ref{fig:power}. While the ``wiggles'' do
not significantly disturb the potential itself, they are relevant for
parameters that depend on derivatives of the potential. The modulations
only slightly affect $r$ (which mainly depends on $\epsilon$ and thus 
the first derivative of the potential), but have an important impact
on $n_s$ (which depends on the second derivative of the potential).
This allows a wider range of values of the spectral index $n_s$ as
compared to simple natural inflation. 

As we have pointed out, the occurrence of subleading instantons in modulated natural inflation ensures
the consistency of the model with the weak gravity conjecture. This 
feature is a direct imprint of the string theory origin of the potential. A more
detailed study about the possibilities to obtain this model in concrete string compactifications
is left for future work. 

\subsection*{Acknowledgments}
This work was supported by the SFB-Transregio TR33 ``The Dark Universe''
(Deutsche Forschungsgemeinschaft).

\bibliography{natural}
\end{document}